\documentstyle [preprint,aps] {revtex}

\begin{document}
\draft

\title{Making predictions in eternally inflating universe}

\author{Alexander Vilenkin}
\address{Institute of Cosmology, Department of Physics and Astronomy,
Tufts University, Medford, MA 02155, USA}

\maketitle

\begin{abstract}
Eternally inflating universes can contain large thermalized regions with
different values of the constants of Nature and with different density
fluctuation spectra.  To find the probability for a `typical' observer to
detect a certain set of constants, or a certain fluctuation spectrum, one needs
to compare the volumes occupied by different types of regions.  If the volumes
are taken on an equal-time hypersurface, the results of such a comparison are
extremely sensitive to the choice of the time variable $t$.  Here, I propose a
method of comparing the volumes which is rather insensitive to the choice of
$t$.  The method is then applied to evaluate the relative probability of
different minima of the inflaton potential and the probability distribution for
the density fluctuation spectra.
\end{abstract}

\narrowtext

\section{Introduction}

Inflation is a state of rapid (quasi-exponential) expansion of the universe
\cite{inflation}.  The inflationary expansion is driven by the potential energy
of a scalar field $\varphi$ (called the `inflaton'), while the field slowly
`rolls down' its potential $V(\varphi)$.  When $\varphi$ reaches the minimum of
the potential, this vacuum energy thermalizes, and inflation is followed by the
usual radiation-dominated expansion.

One of the striking aspects of the inflationary cosmology is that, generically,
inflation never ends \cite{AV83,Linde86}.  The evolution of the field
$\varphi$ is influenced by quantum fluctuations, and as a result thermalization
does not occur simultaneously in different parts of the universe.  Inflating
regions constantly undergo thermalization, but the exponential expansion of the
remaining regions more than compensates for the loss, so that the inflating
volume keeps growing with time.

Eternally inflating universes may contain large thermalized domains with
different values of the parameters determining the low-energy physics.  For
example, if gravity is described by a Brans-Dicke-type theory, different
domains may have different values of the effective gravitational constant
\cite{Garcia94}.  Another possibility is that the potential $V(\varphi)$ has
several minima with different masses and couplings of light particles (and
different symmetry breaking schemes).
Although we cannot observe other regions with different low-energy physics, a
study of their properties and their relative abundances in the Universe
is not necessarily pointless.  If we assume that we are a `typical'
civilization inhabiting the Universe, then such a study may help us understand
why we observe the particular set of low-energy constants in our part of the
universe \cite{Garcia94,AV95,Garcia95}.  (The assumption of being typical was
called the `principle of mediocrity' in Ref.\cite{AV95}.  It is a version of
the `anthropic principle' which has been extensively discussed in the
literature \cite{anthropic}.)

A related question was raised in a very interesting recent paper by A.Linde,
D.Linde and A.Mezhlumian \cite{LLM95} who studied the spectrum of density
fluctuations seen by a `typical' observer in an eternally inflating Universe.
They came to a surprising conclusion that such an observer should find herself
near the center of a very deep minimum of the density field.  If
correct, this result may rule out a wide class of inflationary models.

One could try to implement the mediocrity principle by comparing the number of
civilizations inhabiting regions with different values of the constants (or
different fluctuation spectra) at a given moment of time.  This was the
approach taken by Linde {\it et.al.} \cite{Garcia94,Garcia95,LLM95,LLM94}.
However, one finds that the resulting probability distributions are extremely
sensitive to the choice of the time coordinate $t$ \cite{LLM94,Garcia94}.
Since no particular choice of $t$ appers to be preferred, this casts serious
doubt on any conclusions reached using this approach.  In the present paper I
attempt to define the probability distribution for the constants in a
coordinate-independent way.

The idea is to think in terms of the total number of civilizations in the
entire spacetime, rather than their number on a particular spacelike
hypersurface, $t = {\rm const}$.  To be specific, let us consider `new'
inflation with a potential $V(\varphi)$ of the form illustrated in Fig.1.  The
potential has two minima, and the values $\varphi_*^{(1)}$ and
$\varphi_*^{(2)}$ near the minima correspond to the end of inflation.  The
relative probability for a civilization to be in one type of thermalized region
versus the other can be estimated as \cite{AV95}
\begin{equation}
{{\cal P}^{(1)} \over{{\cal P}^{(2)}}} \sim {{\cal V}_*^{(1)} \nu_{civ}^{(1)}
\over{{\cal V}_*^{(2)} \nu_{civ}^{(2)}}} .
\label{probdef}
\end{equation}
Here, ${\cal V}_*^{(j)}$ is the 3-volume of the hypersurface(s) $\varphi =
\varphi_*^{(j)}$, $j = 1,2$, and $\nu_{civ}^{(j)}$ is the average number of
civilizations originating per unit volume ${\cal V}_*^{(j)}$.  (I assume that
civilizations can originate for only a finite period of time after
thermalization, so that their number per unit thermalized volume is finite).
The dependence of the `human factor' $\nu_{civ}$ on the low-energy physics
constants has been discussed \cite{anthropic} in relation to the anthropic
principle.  Here, I will disregard it and concentrate on determining the
ratio
\begin{equation}
r = {\cal V}_*^{(1)} /{\cal V}_*^{(2)}.
\label{r}
\end{equation}

In models where the Universe is closed and inflation is not eternal, the
volumes ${\cal V}_*^{(j)}$ are finite, and the ratio $r$ is well defined.
However, in an eternally inflating universe, ${\cal V}_*^{(1)}$ and ${\cal
V}_*^{(2)}$ are infinite and have to be regulated.  If one simply cuts them off
by introducing a hypersurface $t = {\rm const}$ and including only parts of the
volumes in the past of that hypersurface, then again one finds \cite{AV95} that
the ratio $r$ is highly sensitive to the choice of the cutoff hypersurface, so
we are back to our original problem.  Here, I would like to suggest an
alternative regularization procedure which appears to give more reasonable
results and which is more robust with respect to changes of the time variable.

The rest of the paper is organized as follows.  The general spacetime structure
of an eternally inflating universe is discussed in the following Section.  The
proposed regularization procedure is outlined in Section 3, and a formalism for
calculating relative probabilities for different minima of $V(\varphi)$, which
is based on this procedure, is introduced in Section 4.  The dependence of the
resulting probabilities on the choice of the time variable is discussed in
Section 5, with a specific example worked out in Section 6.  The spectrum of
density fluctuations detected by a typical observer is calculated in Section 7,
and the result is compared to that of Linde {\it et.al.} \cite{LLM95}.  The
main results of the paper are summarized and discussed in Section 8.

\section{Inflationary spacetimes}

An inflating Universe can be locally described using the synchronous
coordinates,
\begin{equation}
ds^2 = d\tau^2 - a^2 ({\bf x}, \tau) d{\bf x}^2 .
\label{metric}
\end{equation}
The lines of ${\bf x}={\rm const}$ in this metric are timelike geodesics
corresponding to the worldlines of co-moving observers, and the coordinate
system is well defined as long as the geodesics do not cross.  This will start
happening only after thermalization, when matter in some regions will start
collapsing as a result of gravitational instability.  Hence, the synchronous
coordinates (\ref{metric}) can be extended to the future well into the
thermalized region.

The classical evolution of the scale factor $a({\bf x},\tau)$ and of the scalar
field $\varphi ({\bf x},\tau )$ is described by the equations
\begin{equation}
({\dot a}/a)^2 \equiv H^2 \approx 8\pi V(\varphi)/3 ,
\label{H}
\end{equation}
\begin{equation}
{\dot \varphi} \approx -H'(\varphi)/4\pi ,
\label{phieq}
\end{equation}
where dots represent derivatives with respect to $\tau$ and I use Planck units,
${\hbar} = c = G = 1$.  The potential $V(\varphi)$ is assumed to be of the form
shown in Fig.1.  It is assumed also that the field $\varphi$ is a
slowly-varying function of the coordinates, so that spatial gradients of
$\varphi$ can be neglected and ${\dot \varphi}^2 \ll 2V(\varphi)$.  With the
aid of Eqs.(\ref{H}),(\ref{phieq}), the latter condition can be expressed as
\begin{equation}
H' \ll 6H .
\label{slorollcond}
\end{equation}
This condition is violated near the points $\varphi = \varphi_*^{(j)}$,
signalling the end of inflation.  The slow variation of $\varphi$ implies that
$H$ is also a slowly-varying function of ${\bf x}$ and $\tau$, and thus the
spacetime is locally close to DeSitter, with a horizon length $H^{-1}$.

Quantum fluctuations of the field $\varphi$ can be pictured as a `random walk'
[superimposed on the classical motion (\ref{phieq})] in which $\varphi$
undergoes random steps of {\it rms} magnitude $(\delta \varphi)_{\rm
rms}=H/2\pi$ per Hubble time, $\delta\tau = H^{-1}$, independently in each
horizon-size region ($\ell \sim H^{-1}$) \cite{Steinh83}.  The fluctuations are
dynamically unimportant if the classical `velocity' $|{\dot \varphi}|$ is much
greater than the characteristic speed of the random walk, $(\delta\varphi)_{\rm
rms}/\delta\tau = H^2/2\pi$, which gives
\begin{equation}
H' \gg H^2 .
\label{quantcond}
\end{equation}
This condition is violated in the region $\varphi_0^{(1)} < \varphi <
\varphi_0^{(2)}$ near the top of the potential (see Fig.1).  The dynamics of
$\varphi$ in this region is dominated by quantum fluctuations.  The
deterministic slow-roll regions are bounded by the values $\varphi_0^{(j)}$ and
$\varphi_*^{(j)}$.

The spacetime of the inflating universe is schematically represented in Fig.2.
Thermalized regions of different type are shown by different kinds of shading;
their boundaries are the surfaces $\varphi = \varphi_*^{(j)}$.  These
boundaries expand into the inflating region, but become asymptotically static
in the coordinate space as $\tau \to \infty$.  (If the speed of expansion
remained finite at large $\tau$, then thermalized regions would merge and
inflation would terminate everywhere).  This does not mean, however, that the
boundary surfaces become timelike at large $\tau$.  To picture the geometry of
these surfaces, one has to keep in mind that physical lengths in an inflating
universe are related to coordinate differences by an exponentially growing
scale
factor.  As a result, the boundaries of thermalized regions (as well as all
surfaces of constant $\varphi$ in the slow-roll regime) are spacelike, and are
in fact very flat.

The spacelike character of the constant-$\varphi$ surfaces can be understood as
follows \cite{Borde}.  Consider the normal vector to the surfaces,
$\partial_\mu \varphi$.  We have
\begin{equation}
\partial_\mu \varphi \partial^\mu \varphi = {\dot \varphi}^2-a^{-2} ({\bf
\nabla}\varphi )^2 .
\end{equation}
The spatial gradients of $\varphi$ are caused by quantum fluctuations.  On the
scale of the horizon, the gradient is of the order $a^{-1}|{\bf \nabla}\varphi|
\sim (\delta \varphi)_{\rm rms}/H^{-1} \sim H^2 /2\pi$, and is even smaller on
larger scales.  On the other hand, from Eqs.(\ref{phieq}),(\ref{quantcond}),
$|{\dot \varphi}| \gg H^2 /2\pi$.  Hence, $\partial_\mu \varphi \partial^\mu
\varphi > 0$, and the surfaces $\varphi = {\rm const}$ are spacelike.
Moreover, these surfaces are locally nearly parallel to the surfaces $\tau =
{\rm const}$ (which correspond to horizontal lines in Fig.2).  This can be seen
by considering the scalar product of the unit normals, $u^\mu = (1,0,0,0)$ and
$n_\mu = \partial_\mu \varphi (\partial_\nu \varphi \partial^\nu
\varphi)^{-1/2}$,
\begin{equation}
u^\mu n_\mu = [1 - a^{-2}({\bf \nabla}\varphi)^2 /{\dot \varphi}^2 ]^{-1/2}
\approx 1 .
\label{cos}
\end{equation}

For an internal observer in one of the thermalized regions, the surface
$\varphi = \varphi_*$ at the boundary of that region plays the role of the big
bang.  The natural choice of the time coordinate in the vicinity of that
surface is $t = \varphi$, so that the constant-$t$ surfaces are (nearly)
surfaces of constant energy.  Since these surfaces are infinite, the observer
finds herself in an infinite thermalized universe, which is causally
disconnected from the other thermalized regions.  The situation here is similar
to that in the `open-universe' inflation \cite{openinf} where thermalized
regions are located in the interiors of expanding bubbles and have the geometry
of open ($k=-1$) Robertson-Walker universes.

We see that, depending on one's choice of the time coordinate $t$, an
equal-time surface may cross many thermalized regions of different types (e.g.,
for $t=\tau$), may cross no such regions at all (say, for $t=\varphi$ with
$\varphi$ in the slow-roll range), or may lie entirely in a single thermalized
region.  Not surprisingly, with a suitable choice of $t$, one can get any
result for the volume ratio $r$ in Eq.(\ref{r}).

\section{The proposal}

Let us choose a region in which the
field $\varphi$ is close to the top of the potential ($\varphi \approx 0$) at
some moment of time in the coordinates (\ref{metric}).
We can set $\tau
=0$ and $a = 1$ at that moment.  The choice of the initial moment and of the
particular region is
unimportant, due to the self-similar nature of eternal inflation
\cite{Aryal,LLM94}.
Let us now introduce a large number $N$ of co-moving observers
uniformly spread over the region.  Some of these observers will end up in
thermalized regions of type 1 and others in regions of type 2 (see Fig.3).
The initial 3-volume of the region ${\cal V}_0$ has to be sufficiently large,
so that the corresponding co-moving volume includes
a large number of thermalized regions of both types.  (Alternatively,
one could consider an ensemble consisting of a large number of small regions).
The initial volume per observer is ${\cal V}_0 /N$, and if $a_{*i}$ is the
scale
factor for the $i$-th observer at the time when she reaches $\varphi_*^{(j)}$,
then the 3-volumes of the hypersurfaces $\varphi =\varphi_*^{(j)}$
can be represented as
\begin{equation}
{\cal V}_*^{(j)} = {\cal V}_0 N^{-1} \sum_i^{(j)} a_{*i}^3,
\label{Vstar}
\end{equation}
Here, $j=1,2$, and the summation is over all observers
ending up in thermalized regions of type $j$.  [I disregard the factors $(u^\mu
n_\mu)^{-1} \approx 1$; see Eq.(\ref{cos})].

The sums in Eq.(\ref{Vstar})
are, of course, divergent in the limit $N \to \infty$.  My proposal is to
regularize these sums by cutting off a small fraction $\epsilon$ of terms
having the largest scale factors $a_*$, with the same value of $\epsilon$ for
both types of thermalized regions.  The volume ratio $r$ in Eq.(\ref{r}) can
then be evaluated in the limit $\epsilon \to 0$.

A similar regularization can be implemented
using the proper time $\tau$, or some other `time' $t$, as a cutoff variable
instead of the scale factor (that is, discarding a fraction $\epsilon$ of
observers having the largest values of $t$ when they reach $\varphi =
\varphi_*^{(j)}$).  We shall see that the volume ratio $r$ is not very
sensitive to the choice of the cutoff variable.

The reader may be concerned that coordinate-dependence
in this approach is introduced
from the very beginning, when we choose the initial volume ${\cal V}_0$ on an
equal-time surface in the synchronous coordinate system.  The role of this
volume, however, is only to define an appropriate congruence of geodesic
observers.  Geometrically, the construction can be described as follows.
(i) Find a spacetime region where $\varphi \approx 0$.  (ii) Choose a
hypersurface in this region which has approximately constant, small
intrinsic and extrinsic curvature ($^{(3)}R \approx {\rm const} \lesssim H^2$).
(iii) Consider a congruence of
geodesic observers whose worldlines are orthogonal to this hypersurface.

\section{Calculation of the volume ratio}

With the above prescription, the volume ratio $r$ can be calculated in terms of
the probability distribution ${\cal P}(\varphi, t)$ for the inflaton field
$\varphi$.  This distribution satisfies the diffusion equation
\cite{AV83,Starob86}
\begin{equation}
\partial_t{\cal P}=-\partial_\varphi J,
\label{diffeq}
\end{equation}
where the flux $J(\varphi,t)$ is given by
\begin{equation}
J=-(8\pi^2)^{-1} H^{{\alpha \over{2}}+1}\partial_\varphi \left( H^{{\alpha
\over{2}}+1}{\cal P} \right) - (4\pi)^{-1}H^{\alpha -1}H'{\cal P}.
\label{flux}
\end{equation}
The first, `diffusion' term in Eq.(\ref{flux}) represents quantum fluctuations
of the field $\varphi$, while the second term corresponds to the classical
`drift' due to the potential $V(\varphi)$.  The parameter $\alpha$ is defined
so that for $\alpha = 1$ the time variable $t$ is the proper time $\tau$, while
in general $t$ is related to $\tau$ through
\begin{equation}
dt = H^{1-\alpha}d\tau.
\label{alpha}
\end{equation}
In particular, for $\alpha = 0$, $t= \ln a$.  The boundary conditions at
$\varphi = \varphi_*^{(j)}$ are \cite{LLM94}
\begin{equation}
J(\varphi_*)=-(4\pi)^{-1}H^{\alpha -1}H'{\cal P}(\varphi_*).
\label{bc}
\end{equation}
The initial distribution ${\cal P}(\varphi,0)$ is sharply peaked near $\varphi
=0$ and is normalized so that
\begin{equation}
\int_{\varphi_*^{(1)}}^{\varphi_*^{(2)}}{\cal P}(\varphi,0) d\varphi =1.
\label{norm}
\end{equation}

The implementation of the $\epsilon$-cutoff procedure is most straightforward
with the choice of $\alpha =0$.  Then $a=e^t$, and Eq.(\ref{Vstar}) gives
\begin{equation}
{\cal V}_*^{(j)} = {\cal V}_0 \left|
\int_0^{t_c^{(j)}}J(\varphi_*^{(j)},t)e^{3t}dt \right|,
\label{Vstara}
\end{equation}
where the cutoff `times' $t_c^{(j)}$ are determined from
\begin{equation}
\left| \int_{t_c^{(j)}}^\infty J(\varphi_*^{(j)},t)dt \right| = \epsilon \left|
\int_0^\infty J(\varphi_*^{(j)},t)dt \right| \equiv \epsilon p^{(j)}.
\label{cutoff}
\end{equation}
The conservation of probability implies that
\begin{equation}
p^{(1)} + p^{(2)} =1
\end{equation}
Equations (\ref{Vstara}),(\ref{cutoff}) are easily understood if we note that
$|J(\varphi_*^{(j)},t)|dt$ is the fraction of all observers who end inflation
in a region of type $j$ at time $t$ in the interval $dt$.  We note also that
the cutoff times $t_c^{(j)} \to \infty$ as $\epsilon \to 0$.

The solution of Eq.(\ref{diffeq}) can be represented in the form
\begin{equation}
{\cal P}(\varphi,t)=\sum_{n=1}^\infty \psi_n (\varphi) e^{-\gamma_n t},
\end{equation}
where $0<\gamma_1 <\gamma_2 < ...$.  In the limit $t \to \infty$,
\begin{equation}
{\cal P}(\varphi,t \to \infty) = \psi_1(\varphi)e^{-\gamma_1 t}.
\label{larget}
\end{equation}

The eigenfunction $\psi_1 (\varphi)$ can be found explicitely in the slow-roll
regions, away
from the top of the potential, where the diffusion term in Eq.(\ref{flux}) is
negligible.  The equation for $\psi_1 (\varphi)$ in these regions takes the
form
\begin{equation}
\partial_\varphi (H^{-1}H' \psi_1 )=-4\pi \gamma_1 \psi_1 ,
\label{psieq}
\end{equation}
and its solution is
\begin{equation}
\psi_1^{(j)}(\varphi)=c^{(j)}{H(\varphi) \over{H'(\varphi)}}\exp \left[
-4\pi\gamma_1 \int_{\varphi_0^{(j)}}^\varphi {H(\xi) \over{H'(\xi)}}d\xi
\right].
\label{psi}
\end{equation}
The solution $\psi_1^{(j)}(\varphi)$ applies in the range between
$\varphi_0^{(j)}$ and $\varphi_*^{(j)}$.  [Note that Eq.(\ref{psi}) satisfies
the boundary condition (\ref{bc})].  The coefficients $c^{(j)}$ and the
eigenvalue $\gamma_1$ can be found by matching Eq.(\ref{psi}) to the solution
for $\psi_1$ in the range $\varphi_0^{(1)} < \varphi < \varphi_0^{(2)}$.  An
example will be given in Section 6.

We shall now use the asymptotic forms (\ref{larget}),(\ref{psi}) to evaluate
the volume ratio $r$.  For inflation to be eternal, the exponential decay of
the
distribution function (\ref{larget}) should be slower than the expansion of the
inflating regions, $a^3 = e^{3t}$, so that the total inflating volume grows.
Then, $\gamma_1 <3$, and
the integral in Eq.(\ref{Vstara}) for ${\cal V}_*^{(j)}$ is dominated by the
upper limit.  With the aid of Eqs.(\ref{larget}),(\ref{psi}) we can write
\begin{equation}
{\cal V}_*^{(j)} =[4\pi (3-\gamma_1)]^{-1}{\cal V}_0 c^{(j)} \exp \left[
(3-\gamma_1)t_c^{(j)} +\gamma_1t_*^{(j)} \right] ,
\label{Vone}
\end{equation}
where
\begin{equation}
t_*^{(j)}=-4\pi \int_{\varphi_0^{(j)}}^{\varphi_*^{(j)}}{H(\xi) \over{H'(\xi)}}
d\xi
\label{tstar}
\end{equation}
is the `time' of the classical slow roll from $\varphi_0^{(j)}$ to
$\varphi_*^{(j)}$ [see Eqs.(\ref{phieq}),(\ref{alpha})].

The cutoff `times' $t_c^{(j)}$ are determined by Eq.(\ref{cutoff}), and
substituting the asymptotics (\ref{larget}),(\ref{psi}) in the left-hand side
of that equation, we have
\begin{equation}
(4\pi\gamma_1)^{-1}c^{(j)}\exp [\gamma_1 (t_*^{(j)}-t_c^{(j)})]=\epsilon
p^{(j)}.
\label{tc}
\end{equation}
Finally, substituting this into (\ref{Vone}), we obtain
\begin{equation}
{\cal V}_*^{(j)}=c^{(j)}{\cal V}_0 [4\pi (3-\gamma_1 )]^{-1}[4\pi\gamma_1
\epsilon p^{(j)}/c^{(j)}]^{-(3-\gamma_1 )/\gamma_1} \exp [3t_*^{(j)}]
\end{equation}
and
\begin{equation}
r={{\cal V}_*^{(1)} \over{{\cal V}_*^{(2)}}}=\left[{c^{(1)} \over{c^{(2)}}}
\right]^{3/\gamma_1} \left[ {p^{(2)} \over{p^{(1)}}}\right]^{(3-\gamma_1
)/\gamma_1} \left[ {Z_*^{(1)} \over{Z_*^{(2)}}} \right]^3 .
\label{ratio}
\end{equation}
Here,
\begin{equation}
Z_*^{(j)}=\exp \left[ -4\pi \int_{\varphi_0^{(j)}}^{\varphi_*^{(j)}} {H(\xi)
\over{H'(\xi)}}d\xi \right]
\label{Zstar}
\end{equation}
is the expansion factor during the classical slow roll from $\varphi_0^{(j)}$
to $\varphi_*^{(j)}$.

The quantities $\gamma_1$, $c^{(j)}$, $p^{(j)}$, and $Z_*^{(j)}$ all depend on
the shape of the potential $V(\varphi)$.  The eigenvalue $\gamma_1$ and the
ratio $p^{(2)}/p^{(1)}$ are determined mainly by the shape of the potential
near $\varphi =0$, where the dynamics of $\varphi$ is dominated by quantum
fluctuations.  Unless $V(\varphi)$ is very asymmetric near $\varphi =0$, the
numbers of observers ending up in the two types of thermalized regions will
have the same order of magnitude, and thus $p^{(2)}/p^{(1)} \sim 1$.  If the
potential is approximately symmetric near the top, $V(-\varphi ) \approx
V(\varphi)$, then we can set $\varphi_0^{(1)} \approx -\varphi_0^{(2)}$, and
one expects that $\psi_1 (\varphi_0^{(1)}) \sim \psi_1 (\varphi_0^{(2)})$, that
is, $c^{(1)}/c^{(2)} \sim 1$.  In this case, the most sensitive dependence of
$r$ on the shape of $V(\varphi)$ is through the expansion factors $Z_*^{(j)}$,
\begin{equation}
r \sim \left[ Z_*^{(1)} /Z_*^{(2)} \right]^3 .
\label{rest}
\end{equation}

\section{Different time variables}

With a different choice of the time variable ($\alpha \not= 0$), the
calculation of $r$ is somewhat more complicated.  The difficulty is that the
`time'
$t$ is no longer related to the scale factor $a$, and we cannot write the
analogue of of Eq.(\ref{Vstara}) for the volume ${\cal V}_*^{(j)}$.  This can
be done by introducing another distribution function ${\tilde {\cal
P}}(\varphi, t)$, such that ${\tilde {\cal P}}(\varphi,t)d\varphi$ is the
physical volume occupied by regions with $\varphi$ in the interval $d\varphi$
at time $t$.  In other words, ${\tilde {\cal P}}(\varphi,t)d\varphi$ is the
3-volume on a hypersurface $t = {\rm const}$ in which the inflaton field has
values between $\varphi$ and $\varphi + d\varphi$.  The function ${\tilde {\cal
P}}(\varphi,t)$ satisfies the modified diffusion equation
\cite{Goncharov87,Sasaki88,Mijic,LLM94}
\begin{equation}
\partial_t {\tilde {\cal P}}=-\partial_\varphi {\tilde J}+3H^\alpha {\tilde
{\cal P}},
\label{diffeq'}
\end{equation}
where ${\tilde J}$ is given by Eq.(\ref{flux}) with ${\cal P}$ replaced by
${\tilde {\cal P}}$.  At the initial moment, $t=0$, the distributions
${\tilde {\cal P}}$ and ${\cal P}$ are proportional to one another,
\begin{equation}
{\tilde {\cal P}}(\varphi,0) = {\cal V}_0 {\cal P}(\varphi,0).
\end{equation}

The volume in which $\varphi$ rolls down to the value $\varphi_*^{(j)}$ in a
time interval $dt$ is $|{\tilde J}(\varphi_*^{(j)},t)|dt$, and thus the volume
${\cal V}_*^{(j)}$ with a cutoff at $t_c^{(j)}$ can be written as
\begin{equation}
{\cal V}_*^{(j)} = \left| \int_0^{t_c^{(j)}}{\tilde J}(\varphi_*^{(j)},t)dt
\right|.
\label{Vstart}
\end{equation}
The cutoff times $t_c^{(j)}$ are still determined by Eq.(\ref{cutoff}).  Hence,
in order to calculate $r$ for $\alpha \not= 0$, one has to solve both equations
(\ref{diffeq}) and (\ref{diffeq'}) for ${\cal P}$ and ${\tilde {\cal P}}$.

The distribution ${\tilde {\cal P}}(\varphi,t)$ can be expanded in
eigenfunctions,
\begin{equation}
{\tilde {\cal P}}(\varphi,t) = \sum_{n=1}^\infty {\tilde \psi}_n (\varphi)
e^{{\tilde \gamma}_n t},
\end{equation}
where ${\tilde \gamma}_1 > {\tilde \gamma}_2 > ...$.  For inflation to be
eternal, we should have ${\tilde \gamma}_1 >0$.  In the limit $t \to \infty$,
\begin{equation}
{\tilde {\cal P}}(\varphi,t \to \infty)={\tilde \psi}_1 (\varphi) e^{{\tilde
\gamma}_1 t}.
\label{larget'}
\end{equation}

As in the previous Section, the form of ${\tilde \psi}_1 (\varphi)$ in the
slow-roll regime can be found by neglecting the diffusion term in
Eq.(\ref{diffeq'}),
\begin{equation}
\partial_\varphi (H^{\alpha -1}H' {\tilde \psi}_1 )+12\pi H^\alpha {\tilde
\psi}_1 = 4\pi {\tilde \gamma}_1 {\tilde \psi}_1 .
\label{psieq'}
\end{equation}
The solutions of this equation in the regions between $\varphi_0^{(j)}$ and
$\varphi_*^{(j)}$ are
\begin{equation}
{\tilde \psi}_1^{(j)} = {\tilde c}^{(j)}{H^{1-\alpha}(\varphi)
\over{H'(\varphi)}}\exp \left[ -12\pi \int_{\varphi_0^{(j)}}^\varphi {H(\xi)
\over{H'(\xi)}}d\xi +4\pi {\tilde \gamma}_1 \int_{\varphi_0^{(j)}}^\varphi
{H^{1-\alpha}(\xi) \over{H'(\xi)}}d\xi \right],
\label{psi'}
\end{equation}
with ${\tilde c}^{(j)} = {\rm const}$,
and combining Eqs.(\ref{Vstart}),(\ref{larget'}) and (\ref{psi'}), we have
\begin{equation}
{\cal V}_*^{(j)}=[4\pi {\tilde \gamma}_1 ]^{-1}{\tilde c}^{(j)}\left[ Z_*^{(j)}
\right]^3 \exp [{\tilde \gamma}_1 (t_c^{(j)}- t_*^{(j)})].
\label{Vone'}
\end{equation}
Here, $Z_*^{(j)}$ is given by Eq.(\ref{Zstar}) and $t_*^{(j)}$ has the same
meaning as in Eq.(\ref{tstar}) and is given by
\begin{equation}
t_*^{(j)}=-4\pi \int_{\varphi_0^{(j)}}^{\varphi_*^{(j)}}{H^{1-\alpha}(\xi)
\over{H'(\xi)}}d\xi .
\label{tstar'}
\end{equation}
[Note that for $\alpha =0$, Eq.(\ref{psieq'}) reduces to Eq.(\ref{psieq}) with
${\tilde \gamma}_1 =3-\gamma_1$.  Then it is easily understood that
${\tilde c}^{(j)}=
{\cal V}_0 c^{(j)}$, and thus Eqs.(\ref{psi}),(\ref{Vone}) agree with
Eqs.(\ref{psi'}),(\ref{Vone'})].  Now, with the cutoff times $t_c^{(j)}$ from
Eq.(\ref{tc}), we obtain
\begin{equation}
r =  {{\cal V}_*^{(1)} \over{{\cal V}_*^{(2)}}}={{\tilde c}^{(1)} \over{{\tilde
c}^{(2)}}} \left[{c^{(1)}p^{(2)} \over{c^{(2)}p^{(1)}}}\right]^{{\tilde
\gamma}_1 /\gamma_1} \left[{Z_*^{(1)} \over{Z_*^{(2)}}} \right]^3 .
\label{ratio'}
\end{equation}

The coefficients $c^{(j)}, {\tilde c}^{(j)}$ and the eigenvalues $\gamma_1 ,
{\tilde \gamma}_1$ depend on the choice of the parameter $\alpha$, and thus the
ratio $r$ has some dependence on the time parametrization.  [The quantity
$p^{(j)}$ is the fraction of the observers that end up in the $j$-th type of
thermalized regions, and should be independent of $\alpha$].  However, this
dependence appears not to be very strong.  In particular, in cases where the
potential $V(\varphi)$ is nearly symmetric in the region dominated by quantum
fluctuations near the top, we have $c^{(1)}/c^{(2)} \sim {\tilde c}^{(1)}/
{\tilde c}^{(2)} \sim p^{(1)} /p^{(2)} \sim 1$, and the estimate (\ref{rest})
holds for any value of $\alpha$.

In contrast, the result would be extremely sensitive to the choice of $\alpha$
if we used a cutoff at a fixed value of $t$.  Then we would get
\begin{equation}
r=\left|{{\tilde J}(\varphi_*^{(1)},t) \over{{\tilde J}(\varphi_*^{(2)},t)}}
\right| ={{\tilde c}^{(1)} \over{{\tilde c}^{(2)}}} \left[ {Z_*^{(1)}
\over{Z_*^{(2)}}} \right]^3 \exp [-{\tilde \gamma}_1 (t_*^{(1)} -t_*^{(2)} )].
\label{rlinde}
\end{equation}
The quantity $t_*^{(j)}$ is the time it takes the scalar field $\varphi$ to
roll from $\varphi_0^{(j)}$ to $\varphi_*^{(j)}$.  It obviously depends on the
choice of the time coordinate.

To assess the importance of the last factor in (\ref{rlinde}), we need an
estimate for ${\tilde \gamma}_1$.  It can be shown that \cite{footnote1}
\begin{equation}
{\tilde \gamma}_1 \leq 3H_{max}^\alpha ,
\label{gammabound}
\end{equation}
where $H_{max}$ is the largest value of $H(\varphi)$.  [$H_{max} =H(0)$ for the
potential in Fig.1].  For a generic potential,
\begin{equation}
{\tilde \gamma}_1 \sim H_{max}^\alpha
\label{gammaest}
\end{equation}
gives a reasonably good order-of-magnitude estimate.
The quantity
\begin{equation}
d={\tilde \gamma}_1 H_{max}^{-\alpha} <3
\label{fractal}
\end{equation}
can be called the fractal
dimension of the inflating universe \cite{Aryal,footnote2}.
The values of $d \ll 1$
can be obtained only when the potential is fine-tuned so that eternal inflation
is barely possible.

Returning now to Eq.(\ref{rlinde}), we see from
Eqs.(\ref{Zstar}),(\ref{tstar'}), and (\ref{gammaest}) that, if $\alpha >0$,
then ${\tilde \gamma}_1 t_*^{(j)} \geq 3\ln Z_*^{(j)}$, and thus the last
factor in (\ref{rlinde}) is typically no less important than the ratio of the
expansion factors.  In particular, cutoffs at a fixed proper time
($\alpha =1$) and at a fixed scale factor
($\alpha =0$) often give drastically different results.

\section{An example}

The eigenfunctions $\psi_1 (\varphi)$, ${\tilde \psi}_1 (\varphi)$ (and the
corresponding eigenvalues) can be approximately found in the whole range of
$\varphi$ for a potential $V(\varphi)$ shown in Fig.4.  The potential is
completely flat ($V=V_0$) for $\varphi_0^{(1)} < \varphi < \varphi_0^{(2)}$,
satisfies the slow-roll conditions (\ref{slorollcond}),(\ref{quantcond})
outside this range, and is arbitrary otherwise.  Without loss of generality, we
can set $\varphi_0^{(2)} =-\varphi_0^{(1)} \equiv \varphi_0$.

In the flat region of the potential, $|\varphi |<\varphi_0$, the equation for
${\tilde \psi}_1 (\varphi)$ is
\begin{equation}
{{\tilde \psi}_1}'' +\kappa^2 {\tilde \psi}_1 =0 ,
\label{psieqe}
\end{equation}
where
\begin{equation}
\kappa^2 = 8\pi^2 H_0^{-(\alpha +2)} (3H_0^\alpha -{\tilde \gamma}_1 )
\label{kappa}
\end{equation}
and $H_0 =(8\pi V_0 /3)^{1/2}$ is the expansion rate in the flat region.  The
solution of Eq.(\ref{psieqe}) is
\begin{equation}
{\tilde \psi}_1 (\varphi) =A_1 \cos (\kappa \varphi) +A_2 \sin (\kappa\varphi).
\label{psie}
\end{equation}
In the slow-roll regions at $|\varphi| >\varphi_0$ we can use the approximate
solutions (\ref{psi'}).  The matching onditions at $\varphi =\pm\varphi_0$
require that ${\tilde \psi}_1 (\varphi)$ and the corresponding flux ${\tilde
J}_1 (\varphi)$ should be continuous,
\begin{equation}
[{\tilde \psi}_1 ]_{\varphi =\pm \varphi_0} =[{\tilde J}_1 ]_{\varphi =
\pm\varphi_0} =0 .
\label{bce}
\end{equation}
Combining Eqs.(\ref{psi'}),(\ref{psie}) and (\ref{bce}), we obtain the
following system of equations
\begin{equation}
A_1 \cos (\kappa\varphi_0)-A_2 \sin (\kappa\varphi_0)={\tilde
c}^{(1)}H_0^{1-\alpha}/H_1' ,
\end{equation}
\begin{equation}
A_1 \cos (\kappa\varphi_0)+A_2\sin (\kappa\varphi_0)={\tilde
c}^{(2)}H_0^{1-\alpha}/H_2' ,
\end{equation}
\begin{equation}
\kappa H_0^{\alpha +2} [A_1\sin (\kappa\varphi_0)+A_2 \cos (\kappa\varphi_0)] =
2\pi{\tilde c}^{(1)},
\end{equation}
\begin{equation}
\kappa H_0^{\alpha +2}[-A_1 \sin (\kappa\varphi_0)+A_2 \cos (\kappa\varphi_0)]
=2\pi{\tilde c}^{(2)},
\end{equation}
which yields, after some algebra,
\begin{equation}
\beta_1 -\beta_2 +2\kappa\varphi_0 =n\pi ,
\label{kappaeq}
\end{equation}
\begin{equation}
{A_1 \over{A_2}} =\tan \left( {\beta_1 +\beta_2 +n\pi \over{2}}\right) ,
\label{Ae}
\end{equation}
\begin{equation}
{{\tilde c}^{(1)} \over{{\tilde c}^{(2)}}}=(-1)^n{\cos \beta_1 \over{\cos
\beta_2}}.
\label{ce}
\end{equation}
Here, I have introduced the notation
\begin{equation}
\beta_j \equiv \tan^{-1}(\kappa H_0^3 /2\pi H_j' ),
\label{betae}
\end{equation}
where $H_j'$ is the derivative $H'(\varphi)$ taken on the slow-roll side of the
point $\varphi_0^{(j)}$.

Since $H_1' >0$ and $H_2' <0$, the left-hand side of Eq.(\ref{kappaeq}) is a
monotonically growing function of $\kappa$.  This equation, therefore, has a
single solution for any value of $n$.  From Eq.(\ref{kappa}) it is easily
seen that the largest eigenvalue ${\tilde \gamma}_1$ corresponds to the
smallest value of $\kappa^2$, and thus to $n=1$.

For eternal inflation to be possible, we need ${\tilde \gamma}_1 >0$, and
Eq.(\ref{kappa}) yields $\kappa^2 \geq 24\pi^2 H_0^{-2}$.  Together with the
condition (\ref{quantcond}) this implies $|\beta_j | \ll 1$, and from
Eq.(\ref{kappaeq}) with $n=1$ we have
\begin{equation}
\kappa \approx \pi /2\varphi_0.
\label{approxkappa}
\end{equation}
The condition for eternal inflation to be possible in this model can now be
written from Eqs.(\ref{kappa}),(\ref{approxkappa}),
\begin{equation}
\varphi_0 >H_0 /4\sqrt{6} .
\end{equation}
The meaning of this condition is that the size of the flat region $2\varphi_0$
should not be smaller than the size of the random walk `step' $H_0 /2\pi$.

The equations for the function $\psi_1 (\varphi)$ are essentially identical to
those for ${\tilde \psi}_1 (\varphi)$.  Apart from the trivial replacements
${\tilde \psi}_1 \to \psi_1$, ${\tilde c}^{(j)} \to c^{(j)}$, the only change
is that Eq.(\ref{kappa}) is replaced by
\begin{equation}
\kappa^2 =8\pi^2 H_0^{-(\alpha +2)}\gamma_1 .
\label{kappa'}
\end{equation}
Since the smallest eigenvalue $\gamma_1$ again corresponds to the smallest
solution of Eq.(\ref{kappaeq}) for $\kappa$, we conclude that the value of
$\kappa$ is the same for both functions, $\psi_1 (\varphi)$ and ${\tilde
\psi}_1 (\varphi)$.  Hence,
\begin{equation}
{{\tilde \gamma}_1 \over{\gamma_1}}={24\pi^2 \over{H_0^2 \kappa^2}}-1
\label{gammae}
\end{equation}
and
\begin{equation}
c^{(2)}/c^{(1)}={\tilde c}^{(2)}/{\tilde c}^{(1)}.
\label{cce}
\end{equation}

Now we are ready to analyze the dependence of the volume ratio $r$ in
Eq.(\ref{ratio'}) on the parameter $\alpha$.  This parameter does not appear in
Eq.(\ref{kappaeq}) for $\kappa$, and thus $\kappa$ is independent of $\alpha$.
Then it follows immediately from (\ref{gammae}), (\ref{cce}) and (\ref{ce})
that ${\tilde \gamma}_1 /\gamma_1$, ${\tilde c}^{(2)}/{\tilde c}^{(1)}$, and
$c^{(2)}/c^{(1)}$ are also independent of $\alpha$.  Since $p^{(j)}$ and
$Z_*^{(j)}$ are generally $\alpha$-independent,  we conclude that, for the
model considered in this Section, the ratio $r$ is completely insensitive to
the choice of time parametrization.

This conclusion may seem somewhat surprising, but in fact it is not difficult
to understand.  A special property of our model is that there is a one-to-one
correspondence between the time $t$ when a co-moving observer reaches $\varphi
=\varphi_*^{(j)}$ and the expansion factor $a$ along her worldline at that
time,
\begin{equation}
a=\exp [H_0^\alpha (t-t_*^{(j)} )] Z_*^{(j)} .
\end{equation}
For this reason, we can use a time coordinate with any value of $\alpha$ as a
cutoff variable, and the result will be the same as with the scale factor
cutoff.

\section{Density fluctuations}

Let us now consider the spectrum of density perturbations detected by a
`typical' observer.  These perturbations are determined by quantum fluctuations
of the inflaton field $\varphi$ along the observer's worldline \cite{Mukhanov}.
For an ensemble of observers at the same value of $H$, the fluctuations $\delta
\varphi$ are random Gaussian variables with a distribution
\begin{equation}
d{\cal P}_0 (\delta \varphi)=(2\pi\sigma)^{-1/2}\exp\left[ -{(\delta\varphi)^2
\over{2\sigma^2}} \right] d\delta\varphi ,
\label{pzero}
\end{equation}
where
\begin{equation}
\sigma=H/2\pi.
\label{sigma}
\end{equation}
On a given co-moving scale, the fluctuation is produced (that is, it becomes a
part of the classical field) at the time when that length crosses the horizon,
and the value of $H$ in Eq.(\ref{sigma}) should be taken at that moment.
(For the purposes of this discussion, it is sufficient to take a simple-minded
view that the fluctuation occurs instantly).  The gauge-invariant density
fluctuation on the corresponding scale is
\begin{equation}
\delta \rho /\rho \sim 8\pi H\delta\varphi /H' ,
\label{drho}
\end{equation}
where $H' =dH/d\varphi$ and again all quantities are taken at horizon crossing.
Averaging over the distribution (\ref{pzero}) gives the standard result
\cite{Mukhanov}
\begin{equation}
(\delta\rho /\rho)_{rms}\sim 4H^2/|H'|.
\label{drhorms}
\end{equation}

Linde, Linde and Mezhlumian \cite{LLM95} have pointed out that the distribution
(\ref{pzero}) is not necessarily identical to the probability distribution
$d{\cal P}(\delta\varphi)$ for the values of $\delta\varphi$ detected by a
`typical' observer.  There may be a correlation between $\delta\varphi$ and the
expansion factor $a$, in which case observers with different values of
$\delta\varphi$ will contribute to $d{\cal P}(\delta\varphi)$ with different
weights.  Linde {\it et.al.} arrived at a surprising conclusion that `typical'
fluctuations of $\varphi$ can be much greater than the {\it rms} value
(\ref{sigma}).  Here, I shall first calculate $d{\cal P}(\delta\varphi)$ using
the $\epsilon$-cutoff procedure and then compare it with the results of Linde
{\it et.al.}

Fluctuations of $\varphi$ on different length scales are statistically
independent and can be treated separately.  We can therefore concentrate on a
single fluctuation at some value $\varphi =\varphi_1$ (the same for all
observers), disregarding all the rest.  We shall assume that this value is in
the slow-roll regime.

To find the probability distribution for $\delta\varphi$, we start with the
ensemble of co-moving observers defined in Section 3.  We can split it into
sub-ensembles, such that the fluctuation $\delta\varphi$ at $\varphi
=\varphi_1$ is the same with an accuracy $d\delta\varphi$ for all observers in
each sub-ensemble.  The numbers of observers in different sub-ensembles are
proportional to $d{\cal P}_0 (\delta\varphi)$.  The probability $d{\cal
P}(\delta\varphi)$ is then given by
\begin{equation}
d{\cal P}(\delta\varphi) \propto \sum_i a_{*i}^3 ,
\label{pdef}
\end{equation}
where the summation is done over the observers in the corresponding
sub-ensemble.  (Here we consider only observers ending up in the same type of
thermalized regions).  The sum is regularized by discarding a fraction
$\epsilon$ of observers with the largest values of $a_*$.  (A more careful
analysis shows \cite{AVunp} that the same results are obtained if the sum is
cut off using a time variable with an arbitrary value of $\alpha$).

When the field $\varphi$ undergoes a quantum jump $\delta\varphi$, the amount
of proper time $\tau$ necessary to complete the classical rollover to
$\varphi_*$ is changed by
\begin{equation}
\delta\tau =4\pi\delta\varphi /H_1' ,
\end{equation}
where the subscript `$1$' indicates that the corresponding quantity is taken at
$\varphi =\varphi_1$.  As a result, the volume factor $a_*^3$ acquires an
additional factor
\begin{equation}
f(\delta\varphi) \equiv \exp (3H_1 \delta\tau)=\exp [12\pi (H_1 /H_1')
\delta\varphi ].
\label{factor}
\end{equation}
Since this factor is the same for all observers of the sub-ensemble, the cutoff
in the sum (\ref{pdef}) is not affected by the fluctuation.  That is, the
discarded observers are the same as would be discarded for $\delta\varphi =0$.

Apart from the overall number of observers and the magnitude of the fluctuation
at $\varphi =\varphi_1$, different sub-ensembles are statistically equivalent.
Hence, the probability distribution $d{\cal P}(\delta\varphi)$ is proportional
to
\begin{equation}
d{\cal P}(\delta\varphi) \propto d{\cal P}_0 (\delta\varphi) f(\delta\varphi)
\propto \exp \left[ -{2\pi^2 \over{H_1^2}}\left( \delta\varphi -{3H_1^3
\over{\pi H_1'}} \right)^2 \right] d(\delta\varphi).
\label{p}
\end{equation}

The distribution (\ref{p}) describes fluctuations with a non-zero mean value,
\begin{equation}
<\delta\varphi> =3H_1^3 /\pi H_1' .
\label{phibar}
\end{equation}
Deviations from this mean value are still Gaussian, with the dispersion
\begin{equation}
\sigma =[<(\delta\varphi - <\delta\varphi>)^2>]^{1/2} =
H_1 /2\pi .
\end{equation}
According to Eq.(\ref{phibar}), the inflaton tends to fluctuate in the
direction opposite to the classical roll [see Eq.(\ref{phieq})].
This is easy to understand: `backward' fluctuations prolong inflation and
increase the expansion factor.  From Eqs.(\ref{drho}),(\ref{phibar}), the mean
density fluctuation is
\begin{equation}
<\delta\rho_1>/\rho \sim 24 H_1^4 /{H_1'}^2 .
\label{rhobar}
\end{equation}
The average density profile $<\delta\rho ({\bf x})>/\rho$ can be easily
found for a given $H(\varphi)$.

The appearance of a non-trivial average density distribution is an interesting,
and in principle observable, effect.  However, the magnitude of the effect is
hopelessly small.  From Eqs.(\ref{drhorms}) and (\ref{rhobar}) we have
\begin{equation}
<\delta\rho>/(\delta\rho)_{rms} \sim 6H_1^2 /H_1' \sim (\delta\rho
/\rho)_{rms} \ll 1.
\end{equation}
On scales of astrophysical interest, $(\delta\rho /\rho)_{rms} \sim
10^{-5}$, and (assuming the observed density fluctuations are due to inflation)
$<\delta\rho>/\rho \sim 10^{-10}$.

These results are very different from those obtained by Linde {\it et.al.}
\cite{LLM95} who studied the probability distribution for $\delta\varphi$ on
equal-time surfaces, $t={\rm const}$.  Using the proper time $\tau$ as the time
coordinate, they considered an ensemble of observers who reach $\varphi
=\varphi_*$ at a given moment, with the weight $a_*^3$ assigned to each
observer.  They concluded that a typical observer will detect large quantum
jumps of $\varphi$ {\it in} the direction of the classical roll.  This
conclusion can be qualitatively understood as follows.  If I have to reach the
point $\varphi =\varphi_*$ at a given time $\tau$ with the largest possible
scale factor $a_*$, then the winning strategy for me is to spend as much time
as possible at values of $\varphi$ where $V(\varphi)$ is large and the
expansion is fast, and then quickly rush towards $\varphi_*$.

To make this quantitative, it is convenient to consider fluctuations that bring
$\varphi$ {\it to} a given value $\varphi_1$.  Any further fluctuations can be
disregarded due to statistical independence.  The evolution between
$\varphi_1$ and $\varphi_*$ is the same for all observers.  In particular, it
takes the same time and gives the same expansion factor.  However, {\it prior}
to the fluctuation, the observers had different values of $\varphi =\varphi_1
-\delta\varphi$.

It will also be convenient to allow a small interval $dt$ for the time $t_1$ at
which the observers reach $\varphi_*$.  (For greater generality, I am using a
time variable with an arbitrary parameter $\alpha$).  Then we are interested in
the volume-weighted probability for an observer to reach $\varphi =\varphi_1
-\delta\varphi$ within the interval $dt$.  This is given by $|{\tilde
J}(\varphi_1 -\delta\varphi ,t_1 )|dt$, where ${\tilde J}$ is the flux
associated with the distribution ${\tilde {\cal P}}$ introduced in Section 5.
The probability distribution for $\delta\varphi$ is then proportional to
\begin{equation}
d{\cal P}(\delta\varphi) \propto |{\tilde J}(\varphi_1 -\delta\varphi ,t_1 )|d
{\cal P}_0 (\delta\varphi).
\end{equation}
In the slow-roll range of $\varphi$,
\begin{equation}
{\tilde J} \approx -(4\pi )^{-1} H^{\alpha -1}(\varphi)H'(\varphi){\tilde
\psi}_1 (\varphi) \exp ({\tilde \gamma}_1 t),
\end{equation}
and we can use the approximate form of ${\tilde \psi}_1 (\varphi)$,
Eq.(\ref{psi'}).  Keeping only the factors dependent on $\delta\varphi$, we
have
\begin{equation}
d{\cal P}(\delta\varphi) \propto \exp \left\{ -{2\pi^2 \over{H_1^2}} \left[
\delta\varphi - {H_1^3 \over{\pi H_1'}} (3-{\tilde \gamma}_1 H_1^{-\alpha})
\right]^2 \right\} d(\delta\varphi).
\end{equation}
Note the $\alpha$-dependence of the result.

The average value of $\delta\varphi$ can now be written as
\begin{equation}
<\delta\varphi>= {H_1^3 \over{\pi H_1'}} \left[ 3-d \left( {H_{max}
\over{H_1}}\right)^\alpha \right],
\end{equation}
where the fractal dimension $d$ is given by (\ref{fractal}) and $H_{max} =
{max} \{ H(\varphi) \}$.  For $\alpha =0$, which corresponds to $t= \ln a$,
\begin{equation}
<\delta\varphi> |_{\alpha =0} = {(3-d)H_1^3 \over{\pi H_1'}},
\end{equation}
which is the same as our result (\ref{phibar}), apart from the numerical
coefficient.  But for $\alpha =1$, corresponding to the proper time $t=\tau$,
the result is very different.  The expansion rate $H_1$ for the values
of $\varphi$ corresponding to the observable range of length scales is
typically much smaller than the highest expansion rate $H_{max}$, and thus
\begin{equation}
<\delta\varphi>
|_{\alpha =1} \approx -{{\tilde \gamma}_1 H_1^2 \over{\pi H_1'}},
\end{equation}
and
\begin{equation}
|<\delta\varphi>/(\delta\varphi)_{rms} |_{\alpha =1}
=2dH_{max} H_1 /H_1' .
\end{equation}
This is the result of Linde {\it et.al.}  If one assumes that $d \sim 1$ and
$H_{max} \sim
1$, as it is often done in chaotic inflation scenario \cite{LLM94,AV95}, then
it follows from Eq.(\ref{slorollcond}) that $<\delta\varphi> \gg
(\delta\varphi)_{rms}$ and thus $<\delta\rho> \gg (\delta\rho)_{rms}$.

\section{Discussion}

In this paper I have suggested a cutoff procedure which allows one to assign
probabilities to different types of thermalized volumes in an eternally
inflating universe.  The main advantage of this procedure is that it is rather
insensitive to the choice of the time coordinate $t$ which is used to cut off
the infinite volumes.  In some cases there is essentially no $t$-dependence at
all.  Examples are the relative probability for the two minima for a potential
shown in Fig.4 (Section 6), and the probabilities for different fluctuation
spectra (Section 7).

For a potential with multiple minima, we found that the main factor determining
the relative probability of different minima is the slow-roll expansion factor
$Z_*$ (Section 4).  This factor has a very sensitive dependence on the form of
the potential $V(\varphi)$ in the corresponding slow-roll regions, and we
expect the probabilities for different minima to be vastly different.
If one finds that one minimum is more probable than another, this conclusion is
not likely to be changed by a different choice of the time coordinate, even in
cases where the results do show some $t$-dependence.  Since this dependence is
expected to be rather mild, it can affect only rare borderline cases where the
two probabilities are nearly equal.

One can accept this as a genuine uncertainty of the problem and make
predictions only in cases where one minimum is much more probable than the
others.  An alternative attitude is to assert that there is a preferred choice
of the time variable which gives the `correct' probabilities.  If this second
approach is taken, then there is, arguably, a good reason to take $\alpha =0$
($t=\ln a$) as the preferred choice \cite{Starob86,Salopek}.
The only semiclassical variables that can
be used as clocks in an inflating universe are the inflaton field $\varphi$ and
the scale factor $a$ (at least in the simplest models).  We are comparing
volumes of the hypersurfaces $\varphi ={\rm const}$, and thus the only
remaining variable to be used as a clock is $a$.

The cutoff prescription proposed in this paper can also be used to compare
volumes in different, disconnected, eternally inflating universes.  The world
view suggested by quantum cosmology is that such universes may spontaneously
nucleate out of nothing, and that the constants of Nature may take different
values in different universes \cite{quantcos}.  The principle of mediocrity
suggests that we
think of ourselves as a civilization randomly picked out of an infinite number
of civilizations inhabiting this metauniverse.  The probability distribution
for the constants may then be found by comparing the thermalized volumes in
different universes.  [Of course, one also has to compare the `human factors'
$\nu_{civ}$, see Eq.({\ref{probdef})].  The volumes can be calculated along the
same lines as in Section 4, except now the initial conditions are specified on
the initial hypersurface at the `moment of nucleation' and are determined by
solving the Wheeler-DeWitt equation for the wave function of the universe.

An important difference between comparing different minima of the potential and
different universes is that in the latter case the eigenvalues $\gamma_1$ (and
${\tilde \gamma}_1$) are not generally the same for the two universes.  Since
${\cal V}_*^{(j)} \propto \epsilon^{-{\tilde \gamma}_1^{(j)}/\gamma_1^{(j)}}$,
the ratio ${\cal V}_*^{(1)} /{\cal V}_*^{(2)}$ will either vanish or diverge
in the limit
$\epsilon \to 0$, unless ${\tilde \gamma}_1^{(1)}/\gamma_1^{(1)} = {\tilde
\gamma}_1^{(2)}/\gamma_1^{(2)}$.  Hence, only the constants corresponding to
the largest possible value of ${\tilde \gamma}_1 /\gamma_1$ will have a
non-zero probability.  If the condition ${\tilde \gamma}_1 /\gamma_1 ={max}$
does not determine all the constants of Nature, then the probability
distribution for the remaining constants can be found by considering the
$\epsilon$-independent factors in ${\cal V}_*^{(j)}$ (such as $[Z_*^{(j)}]^3$).
These ideas were briefly outlined in Ref.\cite{AV95} and will be discussed in
detail elsewhere.

\acknowledgements
I am grateful to Paul Steinhardt for stimulating correspondence, to Andrei
Linde and Arthur Mezhlumian for their comments on the manuscript, and to
Serge Winitzki for many discussions and several useful
suggestions during the course of this work.  I also acknowledge partial support
by the National Science Foundation.

\begin{figure}

\caption{}{Scalar field potential $V(\varphi)$ with two minima and two
slow-roll regions.  The field values $\varphi_*^{(1)}$ and $\varphi_*^{(2)}$
indicate the end of inflation.  In the region between $\varphi_0^{(1)}$ and
$\varphi_0^{(2)}$ the field dynamics is dominated by quantum fluctuations, and
the slow-roll regions are bounded by $\varphi_0^{(j)}$ and $\varphi_*^{(j)}$.}

\end{figure}

\begin{figure}

\caption{}{A slice through the spacetime of an eternally inflating
universe.  Thermalized regions of two different types are shown by different
shades of grey.}

\end{figure}

\begin{figure}

\caption{}{A congruence of `observers' ending up in different types of
thermalized regions.  The initial volume ${\cal V}_0$ is shown by a horizontal
line, and the observer's wordlines by vertical lines.}

\end{figure}

\begin{figure}

\caption{}{A potential with a flat central regions bordering with two
slow-roll regions.}

\end{figure}


\begin{references}
\bibitem{footnote1}The proof is similar to that for $\gamma_1 >0$
given by Mijic in Ref.\cite{Mijic}.
\bibitem{footnote2}The regions in the vicinity of the surfaces $\varphi = 0$
expand at the highest rate $H_{max}$; their characteristic scale grows as
$R \propto \exp (H_{max}^\alpha t)$.  These regions can be thought of as
the `skeleton' of the inflating universe.  At the same time, the co-moving
volume grows as ${\cal V} \propto \exp ({\tilde \gamma}_1 t) \propto R^d$, with
the `fractal dimension' $d$ given by Eq.(\ref{fractal}).
\end{references}
\end{document}